\documentstyle[epsfig,preprint,aps]{revtex}
\begin{document}
\tighten
\preprint{
\hbox{JLAB-THY-05-301
}}
\draft
\title{Two-scale scalar mesons in nuclei}
%
\author{K. Saito$,^a$\footnote{E-mail: ksaito@ph.noda.tus.ac.jp} 
H. Kouno$,^b$\footnote{kounoh@cc.saga-u.ac.jp} 
K. Tsushima$,^c$\footnote{tsushima@phys.ntu.edu.tw} 
A.W. Thomas$^d$\footnote{awthomas@jlab.org}
}
\address{$^a$Department of Physics, Faculty of Science and Technology \\ 
Tokyo University of Science, Noda 278-8510, Japan\\
$^b$Department of Physics, Saga University, Saga 840-8502, Japan\\
$^c$National Center for Theoretical Sciences at Taipei, Taipei 10617, Taiwan\\
$^d$Thomas Jefferson National Accelerator Facility, USA}
\maketitle
\begin{abstract}
We generalize the linear $\sigma$ model in order to develop 
a chiral-invariant model of nuclear structure.
The model is {\em natural}, 
and contains not only the usual $\sigma$ meson which is 
the chiral partner of the pion but also a new chiral-singlet that is 
responsible for the medium-range nucleon-nucleon attraction. 
This approach  provides significant advantages in terms of its 
description of nuclear matter and finite nuclei in comparison with  
conventional models based on the linear $\sigma$ model. 

\noindent (Keywords: Linear sigma model, Chiral nuclear model, Scalar polarizability)
\end{abstract}

\vspace{1cm}

Phenomenological relativistic field theories can provide a satisfactory 
description of the 
properties of nuclear matter and finite nuclei\cite{qhd,rft}. 
The fundamental importance  
of chiral symmetry means that it must be incorporated into models of this 
kind. As a result, there is a long history of 
attempts to generalize the linear $\sigma$ model (L$\sigma$M) to 
build nuclear models 
with chiral symmetry. 
However, it is well known that the L$\sigma$M, supplemented by 
a repulsive force associated with $\omega$-exchange, 
leads to bifurcations in the 
equation of state (EOS)\cite{bif,boguta} and that the model 
usually gives a very large 
incompressibility, $K$, and, hence, a very stiff EOS. 
The bifurcation problem may be settled (1) by introducing a $\sigma$-$\omega$ coupling in the mean-field 
approximation (MFA)\cite{boguta} and (2) in the one-loop approximation due to a 
modified effective potential\cite{matsui}. 

It is, however, difficult to build relativistic 
mean field phenomenology with manifest chiral symmetry based upon the conventional 
L$\sigma$M, which contains a scalar meson (namely the $\sigma$ meson) playing 
a dual role as the chiral partner of the pion {\em and} the mediator of the medium-range 
nucleon-nucleon (NN) attraction\cite{tang}. 
Chiral perturbation theory actually forbids the 
identification of this NN attraction with the exchange of the $\sigma$. 
In the L$\sigma$M, two critical constraints are imposed: 
one is the ``Mexican hat'' potential, which gives a very strong nonlinearity of the scalar 
field, and the other is the equality between the 
scalar and pion couplings to the nucleon.  The latter 
eventually requires a large value ($\geq 600$MeV) of the $\sigma$ mass and  
leads to unrealistic, 
oscillating nuclear charge densities, level ordering 
and shell closures\cite{tang}. Furthermore, 
as is well known, the predictions of the L$\sigma$M generally involve 
cancellations among several graphs. 
One example is the $\pi N$ scattering amplitude where the $\sigma$ 
exchange combining with the Born term satisfies the soft pion results, 
whose magnitude is too small.

An alternative is to adopt a nonlinear realization of chiral symmetry. To avoid the unnatural cancellations 
of the L$\sigma$M, the $\sigma$ field is eliminated by 
imposing the constraint that the fields lie on the chiral 
circle, $\sigma^2 + {\vec \pi}^2 = f_\pi^2$ 
(with $f_\pi$ the pion decay constant), 
in the nonlinear model (NLM). This is all fine for chiral 
symmetry. However, at mean-field level, it is 
somewhat frustrating for nuclear physics 
because it is not easy to introduce a scalar-isoscalar channel which 
is responsible for the mid-range NN attraction. 
When studying only elementary processes, the 
cancellations are just a matter of taking care 
but when studying the nuclear many-body problem in MFA, 
it is certainly much simpler to introduce a scalar-isoscalar meson explicitly. 

In the usual NLM, the chiral radius is fixed to $f_\pi$ by mere convenience. 
However, nothing prevents us 
from keeping the fluctuation of the chiral circle as a degree 
of freedom, and identifying it 
with the meson which yields the mid-range attraction. 
Chanfray et al.\cite{chanfray} have proposed a 
chiral nuclear model in which the L$\sigma$M is reformulated in 
the standard nonlinear form as far as 
the pion field is concerned but where 
the scalar degree of freedom (called $\theta$), corresponding to  
fluctuations along the chiral radius, is maintained. 
Thus, the model explicitly involves the scalar meson 
for the NN attraction but does not have the unnatural cancellations. 
However, in their calculation 
the mass of the 
$\theta$ meson was predicted to be in the range $0.8-1.0$ GeV, 
which may again be too large to obtain good fits to 
the properties of finite nuclei. 

Combining nonlinear chiral symmetry with the broken scale invariance of QCD is another approach to construct a 
nuclear model\cite{fur}. 
In such a model, the low-energy theorems involving the QCD trace anomaly of the energy-momentum 
tensor are {\em assumed} to be saturated by a scalar 
glueball (gluonium) field with a large (above $1$ GeV) mass 
{\em and} a light scalar field.  Thus, two different scales 
are required in the scalar channel. After 
integrating out the heavy gluonium field, one can construct an 
effective model which contains the chiral-singlet 
scalar field with a mass of about 400-500 MeV.  
Since the light scalar meson can be responsible for 
the mid-range NN attraction, 
the model provides good fits to the bulk properties of finite nuclei and 
single-particle spectra. However, this does not involve the scalar field 
which corresponds to the fluctuation of the 
chiral radius. 

Even in the L$\sigma$M, if a new, light scalar field which 
phenomenologically simulates the mid-range attractive 
NN force is introduced not as the chiral partner of the pion 
but as a chiral singlet, one may be able to 
construct a new chiral-invariant model. 
Here we construct such a model and study the role of scalar mesons in 
symmetric nuclear matter as well as finite nuclei.  As discussed above,  
it may be necessary to include at least two kinds of scalar mesons: 
one the usual $\sigma$ meson with 
a relatively large mass and the other a new, 
chiral-singlet scalar meson with a light mass.  
The $\sigma$ meson is the chiral partner of the pion and 
they form an isoscalar-isovector scalar 
quartet, $(\sigma, {\vec \pi})$.  The fluctuations 
around the stable point on the chiral circle are described by them. 
When the $\sigma$ mass is large, 
the $\sigma$ field itself is relatively reduced in matter and 
it is hence expected that the ``Mexican hat'' potential 
producing the undesirable, strong nonlinearity of 
the $\sigma$ field would also be suppressed.  
In contrast, because of its small mass, the new scalar field is 
enhanced and it may provide the main part of the 
NN attraction.  Our underlying philosophy is that the NN attraction in 
matter is simply dominated by scalar-isoscalar,  
correlated two-pion exchange and that it is this that is being 
simulated by the light scalar meson in MFA. 

Let us first start from the Lagrangian density of the L$\sigma$M with an explicit symmetry breaking term 
\begin{equation}
{\cal L} = {\cal L}_{0} + {\cal L}_{SB} ,  \label{lsm}
\end{equation}
where 
\begin{equation}
{\cal L}_{0} = 
{\bar \psi} [i\gamma^\mu \partial_\mu - g_0 (\sigma + i {\vec \tau}\cdot{\vec \pi}\gamma_5) ] \psi 
+ \frac{1}{2} (\partial_\mu \sigma \partial^\mu \sigma + \partial_\mu {\vec \pi}\cdot \partial^\mu {\vec \pi}) 
-\frac{\lambda}{4} ( \sigma^2 + {\vec \pi}^2 - v^2)^2 ,  \label{l0}
\end{equation}
and ${\cal L}_{SB} = c \sigma$ with $\sigma$ and ${\vec \pi}$ the $\sigma$ and $\pi$ fields, 
respectively.  
Next we introduce a chiral-invariant Lagrangian density for the chiral-singlet scalar meson $s$: 
\begin{equation}
{\cal L}_{s} = \frac{g_s}{f_\pi} {\bar \psi} (\sigma + i {\vec \tau}\cdot{\vec \pi}\gamma_5) F(s) \psi 
+ \frac{1}{2} (\partial_\mu s \partial^\mu s - m_s^2 s^2) ,  \label{ls}
\end{equation}
with $g_s$ the strength of the $s$-N coupling and $m_s$ the $s$-meson mass. 
The first term describes the interaction between the $s$ meson and nucleon, which is similar to that 
adopted in Eq.(20) of Ref.\cite{delorme}. 
In general, $F(s)$ is allowed to be an arbitrary function of the scalar field. 
However, in the present model we choose the simplest one, 
namely, $F(s) = s$. Note that, as demonstrated 
in the QMC model\cite{qmc,qmc2}, this complexity 
is associated with the fact that the nucleon has the internal structure 
which responds to the mean scalar fields. 
In principle one could also multiply this interaction by an arbitrary function 
of $\sigma^2 + {\vec \pi}^2$. 
However, since the $\sigma$ field is relatively suppressed 
because of its large mass, we ignore 
such possibilities in this initial investigation. 
We simply add this Lagrangian density with $F(s) = s$ to the L$\sigma$M. 

It is also vital to add the repulsive NN interaction due to the $\omega$-meson exchange. It can be introduced as a 
gauge-like boson\cite{boguta}, that is, we add the kinetic energy term for the $\omega$ meson and replace 
the derivative by a covariant 
{}form: $\partial_\mu \to D_\mu = \partial_\mu  
+ i g_v \omega_\mu$, with $g_v$ the $\omega$-N coupling constant. 

As usual, we shift the $\sigma$ field ($\sigma \to f_\pi - \sigma$) 
and eliminate $\lambda$, $v$ and $c$ in favor of the free masses of 
the nucleon, $\sigma$ 
and $\pi$ mesons ($M$, $m_\sigma$ and $m_\pi$, respectively).  
(Notice that hereafter $\sigma$ denotes 
the shifted, positive mean value.) 
The free masses are generated by spontaneous 
symmetry breaking. Taking MFA, the total Lagrangian density then reads 
\begin{eqnarray}
{\cal L} &=& 
{\bar \psi} [ i\gamma^\mu \partial_\mu - M^*(\sigma, s) - g_v \gamma_0 \omega ] \psi 
+ \frac{1}{2} (\partial_\mu \sigma \partial^\mu \sigma + \partial_\mu s \partial^\mu s) 
- \frac{1}{2}(m_\sigma^2 \sigma^2 + m_s^2 s^2)  \nonumber \\
&-& \frac{1}{2} \partial_\mu \omega \partial^\mu \omega + \frac{1}{2}m_\omega^{*2}(\sigma, s) \omega^2 
- V(\sigma)  ,  \label{ltot}
\end{eqnarray}
where the pion field vanishes, because the nuclear ground state has good 
parity. The time component 
of the $\omega$ field is simply denoted by $\omega$ 
and the ``Mexican hat'' potential, $V$, is 
\begin{equation}
V(\sigma) = \frac{1-R}{8}\left( \frac{m_\sigma}{f_\pi}\right)^2 \sigma^3 (\sigma -4f_\pi) , \label{mex}
\end{equation}
with $R=(m_\pi/m_\sigma)^2$ and $m_\pi(=138$ MeV) the pion mass. 
In the chiral limit, $R$ becomes $0$. 

The effective nucleon and $\omega$ meson masses are respectively given by 
\begin{eqnarray}
M^*(\sigma, s) &=& (M - g_s s)\left( 1 - \frac{\sigma}{f_\pi} \right) = 
M - g_0 \sigma -g_s s + \frac{1}{M}(g_0 \sigma)(g_s s) , \label{efnm} \\
m_\omega^{*2}(\sigma, s) &=& 
m_\omega^2 \left[\left( 1-\frac{\sigma}{f_\pi}\right)^2 +\left(\frac{s}{f_\pi}\right)^2 \right] . \label{efom}
\end{eqnarray}
Here the free masses and the pion decay constant in vacuum are taken to be  
$M=939$ MeV, $m_\omega =783$ MeV and $f_\pi =93$ MeV. 
Because the pion decay constant is fixed, the coupling constants $g_0$ and $g_v$ are automatically determined 
through the relations $M=f_\pi g_0$ and $m_\omega = g_v f_\pi$. 

This gives the total energy per nucleon (for symmetric nuclear matter) at baryon density $\rho_B$: 
\begin{equation}
\frac{E}{A} = \frac{4}{(2\pi)^3 \rho_B} \int d{\vec k}\ \sqrt{M^{*2} + {\vec k}^2} 
+ \frac{1}{2\rho_B} (m_\sigma^2 \sigma^2 + m_s^2 s^2) + \frac{g_v^2 \rho_B}{2m_\omega^{*2}} 
+\frac{V(\sigma)}{\rho_B}. \label{tote}
\end{equation}
There are three parameters to be determined: the scalar meson masses, $m_\sigma$ and $m_s$, and the 
coupling constant $g_s$.  First, let us try to fix $m_\sigma$ to be a relatively large mass around $1$ GeV. 
The remaining parameters, $m_s$ and $g_s$, are then chosen so as to produce the saturation 
property of symmetric nuclear matter: $E/A -M = -15.7$ MeV at $\rho_0 = 0.17$ fm$^{-3}$ ($\rho_0$ the 
normal nuclear matter density). 

This Lagrangian, however, produces a very small incompressibility, 
around $100$ MeV.  As seen in 
Eq.(\ref{efom}), the coupling of the $\sigma$ meson to the $\omega$ reduces the $\omega$ mass in 
matter, while the light scalar meson enhances it.  Since the $\omega$ mass 
is eventually not reduced much at large $\rho_B$, the 
repulsive force is insufficient and the 
model cannot produce the correct incompressibility ($K = 210 \pm 30$ MeV). 
If the effective $\omega$ mass is reduced significantly, 
the EOS becomes stiff\cite{kouno}. 
To cure this defect, let us add a (chiral-invariant) term 
\begin{equation}
{\cal L}_{new} = - \frac{\zeta}{2f_\pi} g_v^2 \omega ^2 s^3 ,   \label{new}
\end{equation}
to the Lagrangian density.  This term changes the effective $\omega$ meson mass as
\begin{equation}
m_\omega^{*2} = 
m_\omega^2 \left[\left( 1-\frac{\sigma}{f_\pi}\right)^2 + 
\left( 1 -\zeta \frac{s}{f_\pi} \right) \left( \frac{s}{f_\pi} \right)^2 \right] . \label{efomn}
\end{equation}
The new parameter $\zeta$ can control $m_\omega^*$ around $\rho_0$ and 
provide a more realistic incompressibility. Eq.(\ref{efomn}) also implies that, as in the nucleon,  
the $\omega$ meson has internal structure which depends on 
the scalar fields~\cite{qmc2}.  

\begin{figure}
\begin{center}
\epsfig{file=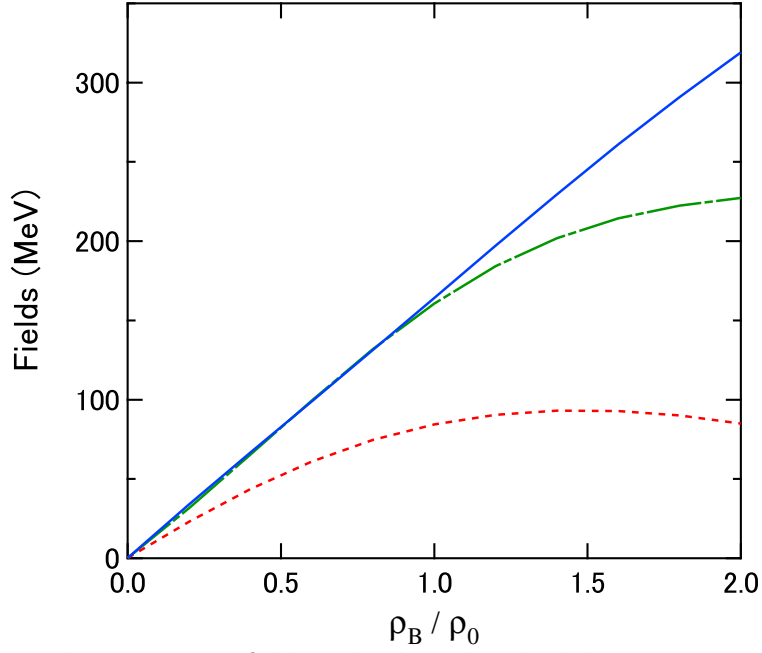,height=9cm}
\caption{
Scalar and vector mean fields.  The dotted and solid curves are, respectively, for the 
$\sigma$ ($g_0 \sigma$) and $s$ ($g_s s$) fields, while the dot-dashed one is for the 
$\omega$ ($g_\omega \omega$) field. The $\sigma$ mass 
is taken to be $1050$ MeV (see Table \protect\ref{tab:cc}). 
}
\label{fig:fields}
\end{center}
\end{figure}

Now we are in a position to show our results for nuclear matter. 
Varying the $\sigma$ 
meson mass, we determine the three parameters, 
$m_s$, $g_s$ and $\zeta$, so as to reproduce 
the saturation condition and the correct $K$.  
In Table~\ref{tab:cc}, we show the nuclear 
matter properties at $\rho_0$.  We also present the scalar and vector 
mean fields as functions of nuclear matter density in Fig.\ref{fig:fields}.  
In our calculation, the $\sigma$ mass varies around $1$ GeV. 
As discussed above, the mass of the light scalar meson 
should be about $400-500$ MeV 
to get good fits to properties of finite nuclei. From 
the table, in the case of $m_\sigma = 
1.0 - 1.05$ GeV, the $s$ meson mass is in the desired range. 
This is not a trivial fact. 
In contrast, in the case where   
$m_\sigma \geq 1.1$ GeV ($m_\sigma \leq 0.95$ GeV), $m_s$ 
seems too large (small).  Furthermore, 
as expected, with increasing $\sigma$ mass, 
the contribution of the $\sigma$ field to the total 
energy is suppressed (see the table and figure) and 
the ``Mexican hat'' potential is correspondingly much reduced.   
Note that, in the chiral model of Boguta\cite{boguta}, we find 
$g_0 \sigma \simeq 200$ MeV and $V/\rho_B \simeq -21$ MeV at $\rho_0$. 

The nucleon mass is reduced by about $25$\% at $\rho_0$. 
In Eq.(\ref{efnm}), the linear terms of the scalar fields 
reduce the mass while the quadratic term increases 
it. It should be emphasized here that the reduction of 
the nucleon mass depends on the $s$ field as well as the 
$\sigma$, where the $\sigma$ may be directly related to the 
change of quark condensate in matter (see 
discussions below), while the $s$-meson contribution 
is related to the mid-range NN attractive force. 
In this model, the two different origins of the mass 
reduction are simultaneously present. 

The contribution of the quadratic term in $M^*$ has already been 
studied in detail in the QMC model from 
the point of view of the quark substructure of the 
nucleon\cite{qmc,qmc2}.  In the QMC model, 
the effective nucleon mass is approximated as 
$M^* \approx M - g_s s + \frac{a}{2}(g_s s)^2$, where 
the coefficient of the quadratic term, $a/2$, 
is known as the scalar polarizability\cite{qmc,tony} 
($a$ is estimated to be $10^{-3}$ MeV$^{-1}$\cite{qmc2}). 
In the present case, since numerically 
$g_0 \sigma \simeq 0.5 \times g_s s$ for the proper $\sigma$ mass 
(see the table \ref{tab:cc}), 
the scalar polarizability is of order
 $1/2M \simeq 0.5 \times 10^{-3}$ MeV$^{-1}$, which is just the same as 
that in the QMC model. 

We now wish to relate 
the change of the nucleon mass to the evolution of the 
quark condensate $\langle {\bar q}q(\rho_B) \rangle$ in 
matter. The quark condensate and its evolution at finite 
density can be obtained by identifying 
the symmetry breaking pieces of QCD with that of our 
Lagrangian: $-2m_q {\bar q}q = c \sigma 
= f_\pi m_\pi^2 \sigma$. This shows that the condensate 
evolution is driven by the mean value of $\sigma$, the 
chiral partner of the pion. From this relation 
and the Gell-Mann-Oakes-Renner relation, we get the 
relative modification of the condensate 
(to the vacuum value) at finite density 
\begin{equation}
\frac{\langle {\bar q}q(\rho_B) \rangle}{\langle {\bar q}q(0) \rangle} = 1 - \frac{\sigma}{f_\pi}. \label{evol1}
\end{equation}
Since both the $\sigma$ and $s$ fields at low density are 
exclusively governed by the scalar 
density of the nucleon field $\langle {\bar \psi}\psi \rangle$
\begin{equation}
\sigma \simeq \frac{g_0}{m_\sigma^2} \langle {\bar \psi}\psi \rangle  \ \ \mbox{and} \ \ 
s \simeq \frac{g_s}{m_s^2} \langle {\bar \psi}\psi \rangle , \label{evol2}
\end{equation}
the $s$ field at low density can be related to the $\sigma$ field 
\begin{equation}
s \simeq \left( \frac{g_s}{g_0} \right) \left( \frac{m_\sigma}{m_s} \right)^2 \sigma . \label{evol3}
\end{equation}
Using Eqs.(\ref{efnm}), (\ref{evol1}) and (\ref{evol3}), 
we can {\em numerically} relate the nucleon 
mass at low density to the condensate evolution
\begin{equation}
\frac{M^*}{M} \simeq \frac{\langle {\bar q}q(\rho_B) \rangle}{\langle {\bar q}q(0) \rangle} 
\left[ 1 - \left( \frac{g_s m_\sigma}{g_0 m_s} \right)^2 
\left( 1 - \frac{\langle {\bar q}q(\rho_B) \rangle}{\langle {\bar q}q(0) \rangle} \right) \right] . 
\label{evol4}
\end{equation}
This rather complicated structure comes from the fact that 
the mass reduction depends two different 
sources, both the quark condensate and the NN attractive force. 

Next let us show some properties of finite nuclei.  
As an illustrative example we consider the case of  
$^{40}$Ca. Firstly, instead of $\rho_0 =0.17$ fm$^{-3}$, 
we adopt the value of 
$\rho_0 = 0.155$ fm$^{-3}$, which is closer to the 
interior density of $^{208}$Pb, and again search 
the parameter set for nuclear matter.  
We then find: $m_\sigma = 0.92$ GeV, $m_s = 404.0$ MeV, 
$g_s^2 = 18.07$ and $\zeta = 0.7$, which reproduces 
the saturation condition with $K = 218$ MeV. 
We next solve a set of coupled nonlinear differential 
equations for a finite nucleus, 
that is derived from the total Lagrangian density 
(see Eqs.(\ref{ltot}) and (\ref{new})).  It may be 
solved by a standard iteration procedure\cite{qmc,qmc2}. 
This leads to a value for the binding energy per nucleon of  
$E_B/A = -8.28$ MeV (the observed value is $-8.45$ MeV).  

In Table~\ref{tab:spectra}, the calculated spectra are summarized. 
Because of the relatively smaller scalar and vector fields in 
the present model than in Quantum Hadrodynamics 
(QHD)\cite{qhd}, the spin-orbit splittings are smaller. 
The good agreement in the binding energy per nucleon comes 
at the expense of a reduction in the spin-orbit force. 
Note that there is a strong correlation between the 
effective nucleon mass and the spin-orbit force. 
It is expected that the inclusion of the exchange contributions 
(Fock terms) will increase the absolute values of the single-particle, 
scalar and vector potentials\cite{Krein}. This increase, together with 
the contribution to the spin-orbit force from the Fock term itself,     
may be expected to improve the spin-orbit splitting in finite nuclei. 
We leave this for a future study. 

\begin{figure}
\begin{center}
\epsfig{file=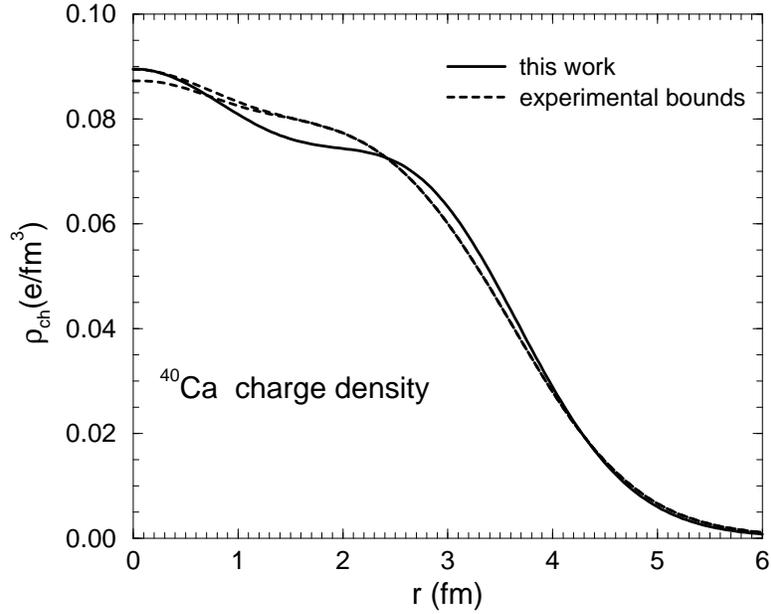,height=10cm,angle=-90}
\caption{
Charge density distribution for $^{40}$Ca. The experimental data are denoted by the dashed 
area\protect\cite{exp1}. 
}
\label{fig:charge}
\end{center}
\end{figure}

The charge density distribution is illustrated in Fig.\ref{fig:charge}.  
Having solved the coupled differential equations, 
we obtain the point-proton and neutron densities in a 
nucleus. It is then necessary to consider the effect of 
the nucleon form factor that gives a 
considerable correction to the density distribution. 
We calculate the charge density by a convolution of the 
point-proton and neutron densities with the proton and 
neutron charge distributions\cite{qmc,formf}. 
We then obtain the root-mean-square (rms) charge radius 
($r_{ch} = 3.44$ fm) and the difference between nuclear 
rms radii for neutrons and protons 
($r_n - r_p = -0.157$ fm) -- the observed values are, 
respectively, $3.48$ fm and $0.05 \pm 0.05$ fm.  

\begin{figure}
\begin{center}
\epsfig{file=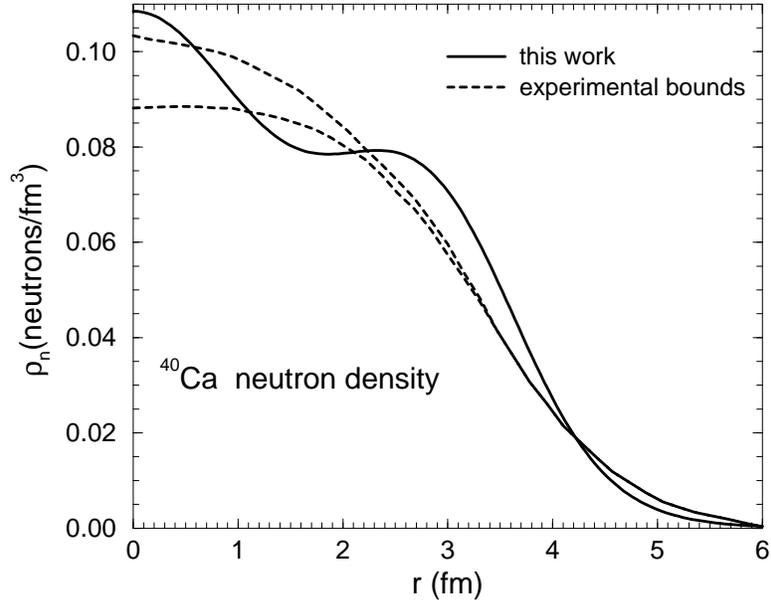,height=10cm,angle=-90}
\caption{
Point-neutron density distribution in $^{40}$Ca. The dashed area represents the empirical fit to 
proton scattering data performed in Ref.\protect\cite{exp2}. 
}
\label{fig:neut}
\end{center}
\end{figure}

As seen in the figure, the calculated charge density distribution 
is close to the experimental 
area. We note that there are no strong oscillations, 
even in the charge density of $^{208}$Pb. 
(In conventional nuclear models based on the L$\sigma$M, 
unrealistic oscillations in the charge density have 
been reported\cite{tang}.)
In Fig.\ref{fig:neut}, we show the neutron density distribution. 
Again we find reasonable agreement with the 
experimental data. The results for the charge and neutron densities 
in our model are very similar to those of 
the other chiral models\cite{fur}. 

In general, an effective field theory at low energy contains an infinite 
number of interaction terms, which incorporate the {\em compositeness} of 
hadrons\cite{qmc2}. It is then expected to involve numerous couplings which may be 
nonrenormalizable. To make sensible calculations, 
Manohar and Georgi\cite{nda} have proposed a systematic way to 
manage such complicated, effective field theories called ``naive
dimensional analysis'' (NDA).
NDA gives rules for assigning a coefficient of the 
appropriate size to any interaction term in an effective Lagrangian. 
After extracting the dimensional factors and 
some appropriate counting factors using NDA, the remaining {\em dimensionless} 
coefficients are all assumed to be of order {\em unity}.  This is the 
so-called {\em naturalness} assumption.  If naturalness is valid, the 
effective Lagrangian can be truncated at a given order with a reasonable 
bound on the truncation error for physical observables.  Then we can control 
the effective Lagrangian, at least at the tree level.  

Here we use NDA to see whether this model gives natural coefficients.
When a generic, interaction Lagrangian density is written as\cite{nat} 
\begin{equation}
{\cal L}_{int.} \sim c_{\ell m n p} \frac{1}{m! n! p!} 
\left( \frac{{\bar \psi} \Gamma (\tau/2) \psi}{f_\pi^2 \Lambda} \right)^{\ell} 
\left( \frac{\sigma}{f_\pi} \right)^m 
\left( \frac{\omega}{f_\pi} \right)^n 
\left( \frac{s}{f_\pi} \right)^p (f_\pi \Lambda)^2 , 
\label{general}
\end{equation}
the overall coupling constant $c_{\ell m n p}$ is 
dimensionless and of ${\cal O}(1)$ if naturalness holds.  
Here $\Gamma$ and $\tau$, respectively, stand for a combination of Dirac matrices and 
isospin operators, and $\Lambda (\sim 1$ GeV) is a large mass scale for the strong interaction.  

The present model has 10 interaction terms 
and the result is shown in Table~\ref{tab:coef}.  
Only three coefficients, $c_{0023}$, $c_{0300}$ and $c_{0400}$, are close to $3$: $c_{0023}$ comes from 
Eq.(\ref{new}) while the latter two are for the "Mexican hat" potential. 
They all are, however, of order unity and almost natural. 
This model is thus {\em natural} as an effective field theory for nuclei. 

In summary, we have developed a chiral nuclear model which is {\em natural},  
based on the L$\sigma$M. 
To describe the properties of 
nuclear matter and finite nuclei at mean-field level, it seems to 
be necessary to introduce two different scales associated with the scalar 
mesons: one is the usual $\sigma$ meson that is 
the chiral partner of the pion, and the other is a new one 
that is treated as a chiral-singlet and 
is responsible for the mid-range NN attraction. 
The present model has three parameters, which are 
determined so as to produce the saturation condition at normal 
nuclear matter density and the proper incompressibility. 
Then, the model automatically predicts two  
scalar scales, that is, one concerns the chiral 
symmetry ($m_\sigma \sim 1$ GeV) and the other relates to the 
NN attraction ($m_s \sim 400-500$ MeV). 
Because the $\sigma$ mass is heavy, 
the contribution of the $\sigma$ field to the total 
energy is relatively suppressed and the ``Mexican hat'' 
potential is very much reduced as well. 
This fact is vital to obtain better fits to the properties of finite nuclei. 
We have demonstrated the single particle spectra, 
charge and neutron densities of 
$^{40}$Ca as well as the properties of symmetric nuclear matter. 
The present approach improves 
conventional nuclear models based on the L$\sigma$M. As a next step, it would be necessary to extend 
this approach to include the $\rho$ meson in order to study unstable nuclei and/or dense matter like 
neutron stars. 

It would be of great interest to study, from the point of view of quark substructure of hadrons, the relation 
between the dependence of hadron masses on the 
scalar fields in this chiral model and that in the 
QMC model\cite{qmc,qmc2}. In particular, if 
the effective $\omega$ mass is sufficiently reduced around 
normal nuclear matter density, it would be possible to form an
$\omega$-nucleus bound state\cite{omega-A} and such 
exotic states may provide significant information on 
chiral symmetry restoration (see Eqs.(\ref{efomn}) and 
(\ref{evol1})).  
{}Finally, we observe that the $\sigma$ mass 
{\em predicted} in this model is very close to the 
scalar glueball mass assumed in Ref.\cite{fur} 
(although it was integrated out and does not appear explicitly). 
Therefore, it would also be very intriguing to study whether there is a 
deeper connection between the $\sigma$ and such 
scalar glueball. 

\vspace{1em}
\noindent
Acknowledgment:
This work was supported by DOE contract DE-AC05-84ER40150,
under which SURA operates Jefferson Laboratory. 

%



%
\begin{table}
\begin{center}
\caption{
Properties of symmetric nuclear matter at normal nuclear matter density ($\rho_0 = 0.17$ fm$^{-3}$). 
The $\sigma$ mass varies in 
the range of $0.9$ and $1.15$ GeV.  Note that, in the present model, $g_0^2 = 101.9$ and $g_v^2 = 70.89$, 
which together with the pion decay constant ($f_\pi =93$ MeV) give the free nucleon and $\omega$ meson masses. 
We take $m_\pi = 138$ MeV. 
All masses, energies and incompressibility are quoted in MeV.
}
\label{tab:cc}
\begin{tabular}[t]{ccccccccc}
$m_\sigma$&$m_s$&$g_s^2$&$\zeta$&$g_0 \sigma$&$g_s s$&$V(\sigma)/\rho_0$&$M^*/M$&$K$\\
\hline 
900  & 346.5 & 12.28 & 0.7  & 130 & 130 &  -6.69 & 0.743 & 214 \\
950 & 388.0 & 17.29 & 0.75 & 111 & 143 &  -4.72 & 0.747 & 216 \\
1000  & 441.7 & 24.48 & 0.8  & 96.4 & 155 &  -3.42 & 0.750 & 214 \\
1050 & 510.7 & 35.00 & 0.85 & 84.4 & 164 &  -2.55 & 0.751 & 214 \\
1100  & 612.1 & 53.08 & 0.85 & 74.6 & 172 &  -1.93 & 0.752 & 212 \\
1150 & 741.9 & 81.35 & 0.85 & 66.5 & 179 &  -1.50 & 0.752 & 218 \\
\end{tabular}
\end{center}
\end{table}
\begin{table}
\begin{center}
\caption{
Single particle energies for $^{40}$Ca. 
All energies are quoted in MeV.
}
\label{tab:spectra}
\begin{tabular}[t]{ccccccc}
 &$1s_{1/2}$&$1p_{3/2}$&$1p_{1/2}$&$1d_{5/2}$&$1d_{3/2}$&$2s_{1/2}$\\
\hline 
proton  & -35.8 & -24.8 & -23.2 & -12.7 & -10.0 & -7.7 \\
neutron & -44.8 & -33.2 & -31.7 & -20.5 & -17.9 & -15.8 \\
\end{tabular}
\end{center}
\end{table}
\begin{table}
\begin{center}
\caption{Interaction terms and corresponding (dimensionless) coupling 
constants for $m_\sigma = 1050$ MeV (see Table\protect\ref{tab:cc}). }
\label{tab:coef}
\begin{tabular}[t]{ccc}
term & $c_{\ell m n p}$ & value \\
\hline
${\bar \psi} \sigma \psi$ & $c_{1100}$ & 0.94 \\
${\bar \psi} s \psi$ & $c_{1001}$ & 0.55 \\
${\bar \psi} \sigma s \psi$ & $c_{1101}$ & 0.55 \\
${\bar \psi} \gamma_0 \omega \psi$ & $c_{1010}$ & 0.78 \\
$ \sigma \omega^2 $ & $c_{0120}$ & 1.2 \\
$ \sigma^2 \omega^2 $ & $c_{0220}$ & 1.2 \\
$ s^2 \omega^2 $ & $c_{0022}$ & 1.2 \\
$ s^3 \omega^2 $ & $c_{0023}$ & 3.1 \\
$ \sigma^3 $ & $c_{0300}$ & 3.3 \\
$ \sigma^4 $ & $c_{0400}$ & 3.3 \\
\end{tabular}
\end{center}
\end{table}
%



\newpage

\end{document}